\documentclass{osa-article}

\journal{osajournal}
\articletype{Research Article}
\usepackage{braket}
\usepackage{esvect}
\usepackage{amsmath}
\usepackage{mathtools}
\newcommand{\AH}[1]{{\color{black}#1}}
\newcommand{\IZ}[1]{{\color{black}#1}}

\graphicspath{{figures/}}
\begin{document}

\title{Parallelization of frequency domain quantum gates: manipulation and distribution of frequency-entangled photon pairs generated by a 21 GHz silicon micro-resonator }

\author{Antoine Henry\authormark{1,2*}, Dario Fioretto\authormark{2}, Lorenzo M. Procopio\authormark{3}, Stéphane Monfray\authormark{4}, Frédéric Boeuf\authormark{4}, Laurent Vivien\authormark{2}, Eric Cassan\authormark{2}, Carlos Ramos\authormark{2}, Kamel Bencheikh\authormark{2}, Isabelle Zaquine\authormark{1}, Nadia Belabas\authormark{2*} }

\address{\authormark{1}LTCI, Télécom Paris, Institut Polytechnique de Paris, 19 place Marguerite Perey, 91120, Palaiseau, France\\
\authormark{2}Université Paris-Saclay, CNRS,Centre for Nanosciences and Nanotechnology, UMR 9001, 10 Boulevard Thomas Gobert, 91120, Palaiseau, France\\
\authormark{3} Weizmann Institute of Science, Rehovot 7610001, Israel\\
\authormark{4}STMicroelectronics SAS, 850 rue Jean Monnet, 38920 Crolles, France}

\email{\authormark{*}nadia.belabas@universite-paris-saclay.fr}
\email{\authormark{*}antoine.henry@telecom-paris.fr}%% email address is required


%%%%%%%%%%%%%%%%%%%%%
%%%%%%%%%%%%%%%%%%
%    ABSTRACT %%%%%%%%
\begin{abstract}

Harnessing the frequency dimension in integrated photonics offers key advantages in terms of scalability, noise resilience, parallelization and compatibility with telecom multiplexing techniques. Integrated ring resonators have been used to generate frequency-entangled states through spontaneous four-wave-mixing. However, state-of-the-art integrated resonators are limited by trade-offs in size, number of frequency modes and spectral separation. We have developed silicon ring resonators with a foot-print below 0.05 mm2 providing more than 70 frequency channels separated by 21 GHz. We exploit the narrow frequency separation to parallelize and independently control 34 single qubit-gates with off-the-shelf electro-optic devices. This allows to fully characterize 17 frequency-bin maximally-entangled qubit pairs by performing quantum state tomography.  We demonstrate for the first time a fully connected 5-user quantum network in the frequency domain. These results are a step towards a new generation of quantum circuits implemented with scalable silicon photonics technology, for applications in quantum computing and secure communications. 
\end{abstract}

\section{Introduction}

%%%%%%%%%%%%%
Frequency encoding provides a resource-efficient way to access a high-dimensional Hilbert space within a single spatial mode, opening the way for scalable quantum information processing. In this scheme, photons can be generated in a superposition of different frequency modes. The superpositions are very robust against phase noise in long-distance propagation. Photon pairs can be generated through non-linear interaction such as spontaneous parametric down conversion (SPDC) or spontaneous four wave mixing (SFWM).
%can create frequency entangled states, where t
The frequency bins are created 
%either 
by external filtering of a wideband nonlinear source \cite{lu_quantum_2018,imany_frequency-domain_2018,ponce_unlocking_2022,khodadad_kashi_spectral_2021}, or by exploiting the inherently discrete frequencies of a resonator \cite{kues_-chip_2017,lu_full_2021,clementi_programmable_2023,borghi_reconfigurable_2023}.

The silicon-on-insulator (SOI) technology provides key advantages for the generation of correlated and entangled photon pairs, including the scalability and the availability of a wide library of high-performance optic and optoelectronic devices. Indeed, silicon photonics has been identified as an enabling technology for quantum information\cite{wang_integrated_2020}. In particular, it is possible to generate bi-photon frequency combs through spontaneous four wave mixing (SFWM) in high-quality-factor microresonators \cite{zhang_correlated_2016,samara_high-rate_2019,yin_frequency_2021,clemmen_continuous_2009}

Moreover in the telecom wavelength range, where the SOI is transparent and efficient for SFWM, off-the-shelf filters, demultiplexers and modulators are available and long-distance interaction between future quantum processors or communication nodes can be achieved using existing classical telecom infrastructures.

%The manipulation of such frequency qubits can be based either on non-linear processes \cite{kobayashi_frequency-domain_2016,raymer_interference_2010}, with spectral modes separated by several hundreds of GHz.
The frequency bins can be manipulated using  non-linear processes like optical frequency conversion\cite{kobayashi_frequency-domain_2016,raymer_interference_2010}.
However, this approach may be hampered by limited configurability and by the optical noise from the pump required to generate the nonlinear phenomena. When generated in the telecom wavelengths, near 1550 nm, frequency bins can be manipulated using off-the-shelf telecom devices such as electro-optic phase modulators (EOM)\cite{olislager_creating_2014,olislager_frequency-bin_2010,bloch_frequency-coded_2007} and programmable filters (PF) \cite{lukens_frequency-encoded_2017,kues_-chip_2017,ponce_unlocking_2022,clementi_programmable_2023}.
%\AH{Frequency-coded quantum manipulation were investigated as early as 2007 in proof of concept studies with attenuated lasers and electro-optic phase modulator \cite{bloch_frequency-coded_2007}. This was followed by \cite{olislager_frequency-bin_2010}, which used frequency-entangled photons generated from a PPLN crystal to measure two-photon correlations and deduce Bell parameters.}
It has been shown that the combination of two EOMs and a PF allows arbitrary qubit transformation if the qubit mode spacing $\Delta f$ is equal to the Radio Frequency $\Omega$ (RF) driving the EOMs \cite{lu_electro-optic_2018,lu_fully_2020}. The reconfigurability of such quantum frequency processors allows for various applications\cite{lu_controlled-not_2019,lu_subatomic_2019,lu_high-dimensional_2022}. 

%Yet, achieving such flexible operation requires driving the EOM with a frequency $\Omega_m$, equal to the spectral separation among adjacent qubit frequency modes $\Delta f$.
The limited bandwidth of standard telecom EOMs limits the maximum qubit frequency spacing, $\Delta f$, to a few tens of gigahertz. In addition, low $\Delta f$ boosts the spectral efficiency, i. e. maximizes the achievable Hilbert dimension for a given available source bandwidth. Yet, achieving such a narrow frequency spacing with integrated resonators is a challenging task.    

Spectral separation of 40-50 GHz has been achieved using silicon nitride ring resonators with a radius near 500 um, yielding a foot-print exceeding 1 mm$^2$
\cite{kues_-chip_2017,imany_50-ghz-spaced_2018,lu_full_2021}. Frequency separation of only $\Delta f$ $\approx$ 20 GHz has been recently achieved by combining silicon rings with resonances separated by 200 GHz \cite{clementi_programmable_2023,borghi_reconfigurable_2023}. This clever design nevertheless gives rise to variable qubit mode spacing and limits the number of achievable frequency modes.  
% It is possible to . Low FSR allowing to drive modulators with a frequency $\Omega_m=FSR$
% However, $\Delta f$ must be compatible with the RF frequency for easier manipulation. While manipulation can be done fore 
% Manipulation of frequ
% In \cite{clementi_programmable_2023, borghi_reconfigurable_2023}, N-dimensional (up to N 
% = 4) frequency-entangled quantum states are generated with FSR $\approx$ 20 GHz, by cascading N rings and thermally tuning their central frequencies. In \cite{lu_full_2021}, manipulation of up to 8-dimensional states is reported, using sources based on both lithium niobate and silicon nitride ring resonators, with a FSR$\approx$40.5 GHz. 

The generated frequency-entangled quantum states can be characterized with electro-optic modulators and programmable filters. A single modulator does not allow for a unitary control of a photonic qubit, but permits the quantum state tomography of high dimensional states \cite{kues_-chip_2017, imany_50-ghz-spaced_2018, lu_full_2021, clementi_programmable_2023, borghi_reconfigurable_2023, ponce_unlocking_2022}. Quantum state tomography can be performed with parallelizable unitary operation on entangled qubits if the spacing is lower than the RF bandwidth. In \cite{lu_quantum_2018}, frequency-entangled qubits from a periodically poled lithium niobate SPDC source filtered by a fibered etalon frequency comb with a spacing of 25 GHz and quantum state tomography is performed with the [EOM-PF-EOM] configuration. The parallelization of two gates allowed the control of two independent frequency qubits.

% The characterization of these entangled state has been done using the electro-optic approach, by performing quantum state tomography. This measurement can be done using a single EOM

% paragraphe à revoir\AH{
% Manipulation of these quantum states has been achieved using a single EOM \cite{kues_-chip_2017, imany_50-ghz-spaced_2018, lu_full_2021, clementi_programmable_2023, borghi_reconfigurable_2023, ponce_unlocking_2022} by performing quantum state tomography or observing Bell-like quantum interferences for up to 7 dimensions. In \cite{lu_quantum_2018}, quantum state tomography is performed with a combination of two EOMs and one PF, in the [EOM-PF-EOM]
% configuration, parallelizing two frequency-domain quantum gates on the signal and idler qubits.}

%Demultiplexing is a technique that can be used to generate several correlated photon pair sources from a broadband source. This approach allows for the entangled state to be encoded in another degree of freedom, such as polarization or time. Distributing these demultiplexed photon pairs can enable the implementation of fully-connected quantum networks\cite{joshi_trusted-node-free_2020,appas_flexible_2021-1,fitzke_scalable_2022,wengerowsky_entanglement-based_2018}.

In this paper, we report the parallelization of 34 tunable electro-optic frequency domain quantum gates, all implemented with a single [EOM-PF-EOM] configuration. 
% In this paper, we report the parallelization of 34 electro-optics frequency domain quantum gates that can be tuned from Identity to a Hadamard transformation and that are implemented with a single [EOM-PF-EOM] device configuration.
% We design to this end a SOI spiral micro-resonator with a FSR of 21 GHz compatible with the bandwidth of the PF and EOM
% We manipulate frequency bins generated through spontaneous four wave mixing in a 
We develop to this end a SOI spiral ring-resonator with a foot-print below 0.05 mm$^2$ and a frequency channel separation 
$\Delta f$ = 21 GHz.
The rings are fabricated using STMicroelectronics’ silicon photonics R\&D and manufacturing platform based on 300-mm SOI wafers and 193-nm-deep-ultraviolet (DUV) lithography, ensuring compatibility with large-scale production\cite{Boeuf:16}.
The narrow spectral separation allows photon pair generation through SFWM on more than 70 frequency modes over a 1.4 THz. In addition, the 21 GHz spectral separation allows implementing parallel and arbitrary qubit transformations, based on [EOM-PF-EOM] scheme implemented with commercially available electro-optic devices. We perform quantum state tomography on frequency domain maximally entangled photons. Based on this approach we perform a proof-of-concept-demonstration of a fully-connected quantum network, where 5 users share a secure key with every other user, using frequency-bin entangled qubits generated by our broadband silicon photon pair source and controlled by 34 parallel tunable electro-optic quantum gates.

%%%%%%%%%
%%%%%      21 GHz Sample
%%%%%###########################################################
\section{21 GHz Silicon On Insulator spiral micro-resonator}

%The micro-resonator used in this experiment is manifactured using large-scale, highly reproducible 300 mm semi-conductor technology by Femot ST. The waveguide and resonators are 900 nm thick.
Our photon pairs are generated through SFWM in a SOI micro-resonator (MR).

The integrated photonic devices were fabricated using STMicroelectronics’ silicon photonics R\&D and manufacturing platform based on 300 mm SOI wafers. The structures were defined with DUV lithography and transferred to the silicon layer with reactive ion etching\cite{Boeuf:16}. A 2 mm-thick PMMA layer was deposited over the chip surface for protection. The thickness of the guiding silicon layer is 300 nm. A waveguide width of 700 nm is chosen to yield small anomalous dispersion near 1550 nm wavelength. 
The ring is shaped as spiral to reduce the footprint of the device. The spiral waveguide length is set to 3.54 mm to yield a free-spectral-range (FSR) near 21 GHz, which determines the frequency channel separation ($\Delta f$).
The waveguide bendings follow a Bezier trajectory to minimize losses due to mode mismatch between straight and curved waveguides\cite{do_wideband_2020}. The spiral resonator has a size of 165 $\mu$m by 255 $\mu$m (<0.05 mm$^2$).
The sample temperature is tuned and stabilized by a Peltier module at  25°$\pm0.01$C.  Cleaved SMF28 fibers, set at a 15° incidence angle, are used to couple light in and out of the chip, through single-etch grating couplers. The measured "fiber-to-fiber" insertion loss is 7.6 dB, i. e. 3.8 dB per coupler including propagation loss. An experimental value of $\text{FSR}=21.18 \pm 0.85$ GHz is obtained by measuring the frequency spacing of adjacent resonances from 1526.7 nm to 1565 nm. The resonator modes exhibit a full width-at-half-maximum (FWHM) of 600 MHz, leading to a quality factor $Q\simeq 3.10^5$.
%The photon pairs are generated through spontaneous four-wave-mixing (SFWM).
We tune the optical pump frequency $\omega_p$ on a resonator mode so
that signal and idler photons are emitted on symmetric resonances at frequency $\frac{\omega_p}{2\pi} \pm n\cdot \text{FSR}$ ($n\in \mathbb{N}$). \IZ{As a result, the measured FSR varies less than 20 MHz over the whole measured SFWM spectrum where the correlated pairs can be observed from 1526.7 to 1553 nm.
The quantum state $\ket{\phi}$ of the generated photon pairs is therefore}

% with a measured dispersion of 2,73$\pm$0.02 MHz/nm.

%The frequencies of the generated signal and idler photons are distributed symmetrically to the pump frequency. The emission spectrum is convoluted by the cavity spectrum, and we adjust the pump frequency $\omega_p$ to be at a resonator mode so that signal and idler photons are emitted respectively at $\omega_p \pm n\cdot \text{FSR}$ ($n\in \mathbb{N}$).

\begin{equation}
    \ket{\phi} = \frac{1}{\sqrt{N}}\sum_{n=1}^{N}e^{i\alpha_n}\ket{I_{n}}\ket{S_{n}},
    \label{state}
\end{equation}
where $I_{n}$ and $S_{n}$ stand for idler and signal frequencies with $I_{n}=\frac{\omega_p}{2\pi}-n\cdot\text{FSR}$ and $S_{n}=\frac{\omega_p}{2\pi}+n\cdot\text{FSR}$, respectively. $\alpha_n$ corresponds to the bi-photon residual spectral phase\cite{lu_full_2021},
% and $\ket{I_n}\ket{S_n}$ is a signal-idler pair where each photon is emited at the frequency $I_n$,$S_n$, 
within the 600 MHz linewidth of the resonances. The ring resonator intrinsically produces frequency-bin entangled qudits of dimension $N$ where $N$ is limited
here %The highest N reachable is determined 
by the bandwidth of the programmable filters (5 THz) and phase matching conditions.

\begin{figure}[h!]
    \centering
    \includegraphics[width=8cm]{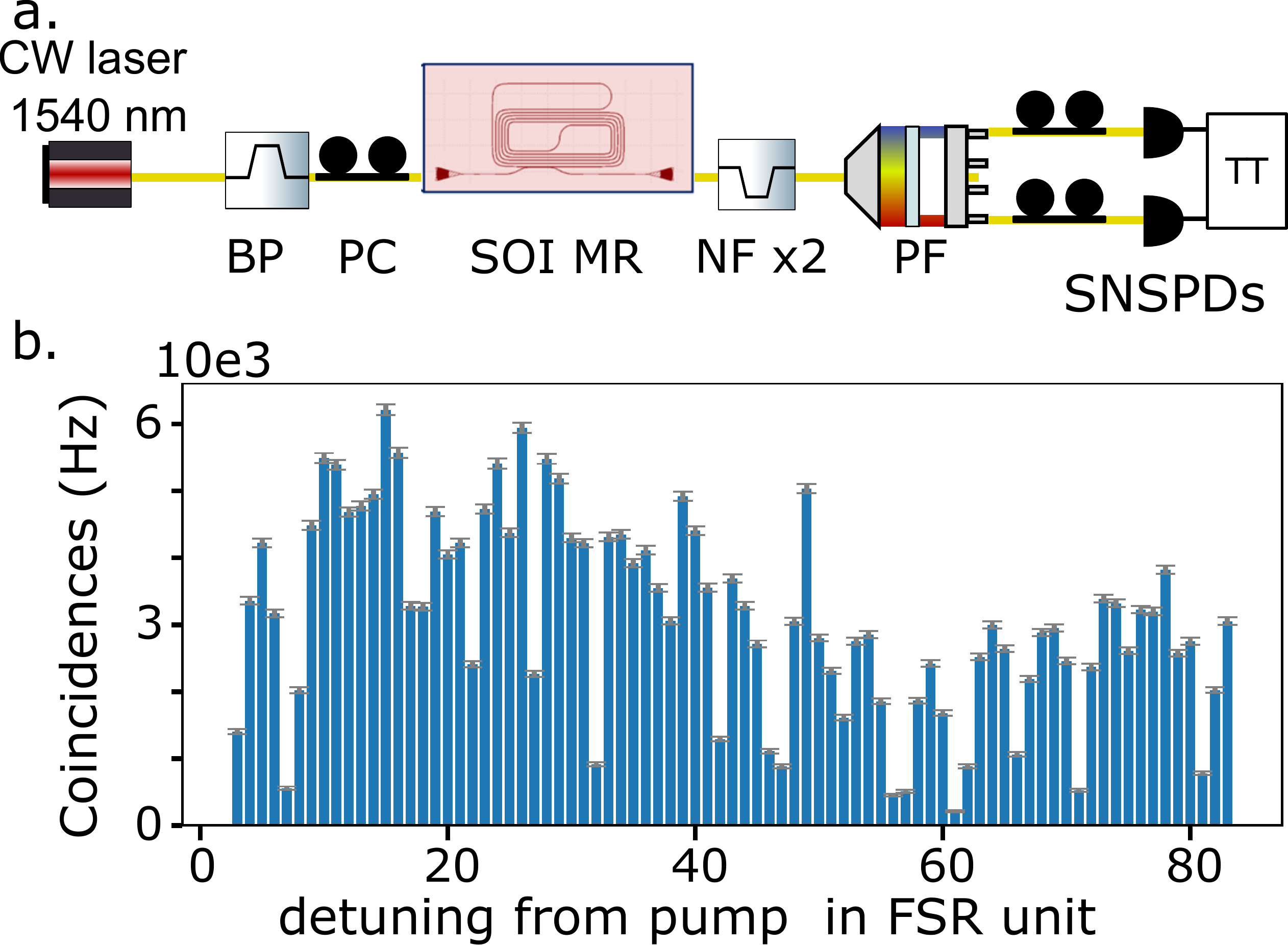}
    \caption{a. Set up for measurement of the Joint Spectral Intensity. BP: Band-pass filter, NF: Notch filter, PF: programmable filter, PC polarization controller, SNSPDs: superconducting single-photon detectors, b. Joint spectral intensity measurement for every accessible signal-idler pair \IZ{from $n=3$ to $n=83$}.}
    \label{fig:JSI}
\end{figure}
To characterize the spectral correlation, we measure the Joint Spectral Intensity of the biphoton state. We use the setup
shown in Fig. \ref{fig:JSI}.a. A bandpass filter (BP) is used to filter out the amplified spontaneous emission of the CW pump laser at 1540 nm up to 40 dB. Two fibered Bragg notch filters with a bandwidth of 80 GHz are used at the output of the SOI MR to filter out 70 dB of the laser light and an additional filtering of the laser light is done with the programmable filter, ensuring  \IZ{more than} 100 dB pump rejection. The spectrally correlated photon pairs are spatially separated by the same programmable filter, sent to Superconducting Nanowire Single Photon Detectors (SNSPDs) with a 70 \% quantum efficiency, and analyzed using the Swabian TimeTagger Ultra (TT). In this paper, the coincidence are taken in a 1ns windows. The dead time of the detectors is $\approx$20 ns, allowing maximum %quantum
efficiency for the range of count rates observed in this work (in the order of 40 kHz). The combined time jitter of the SNSPDs and the TT is around 120 ps.

In Fig.\ref{fig:JSI}.b., we plot the number of coincidences between the signal and idler paths as a function of the selected resonance number $n$ associated with the frequency pairs $I_n,S_n$.
%\ket{I_n}\ket{S_n}$., showing the 
\IZ{This spectral distribution of the photon pairs corresponds to the diagonal elements of the Joint Spectral Intensity.} \AH{We deduce an internal brightness of 15.87$\pm 0.03 \times 10^6$ pairs/s, and measure an heralded $g^{(2)}(0)=0.057\pm 0.007$, for a on-chip power of 0.75 mW.}
%The dimensionality of the generated state is limited by the number of resonances within the total bandwidth of the PF (5 THz).
The low FSR of our resonator allows access to a larger number of resonances compared to previous works \cite{lu_full_2021,imany_50-ghz-spaced_2018,kues_-chip_2017}, providing \IZ{possibilities of parallelization of operations over a larger number of qubits} and perspectives for the processing of higher dimensional quantum states.
%, or parallelization over a larger number of qubits.  
%The dips in the spectral distribution of Fig. \ref{fig:JSI}.b. arise from imperfections in the waveguide of the relatively large resonator required for the small mode spacing we are using.
\IZ{The long cavity nonetheless leads to more propagation losses.}
%and larger rings impact of imperfections in the waveguide, This explains the disparities in the number of coincidences observed for the various signal-idler pairs}.
The overall decrease of the coincidences with increasing $n$ is linked to the spectral transmission of the grating couplers.

%%%%%%%%%%%%%%%%%%%%%%%%%%%%%%%%%%%%%%%%%%%%%%%%%%%%%%%%%%
\section{Frequency-domain Quantum State Tomography}\label{QST_section}

\begin{figure}[h!]
\centering\includegraphics[width=8cm]{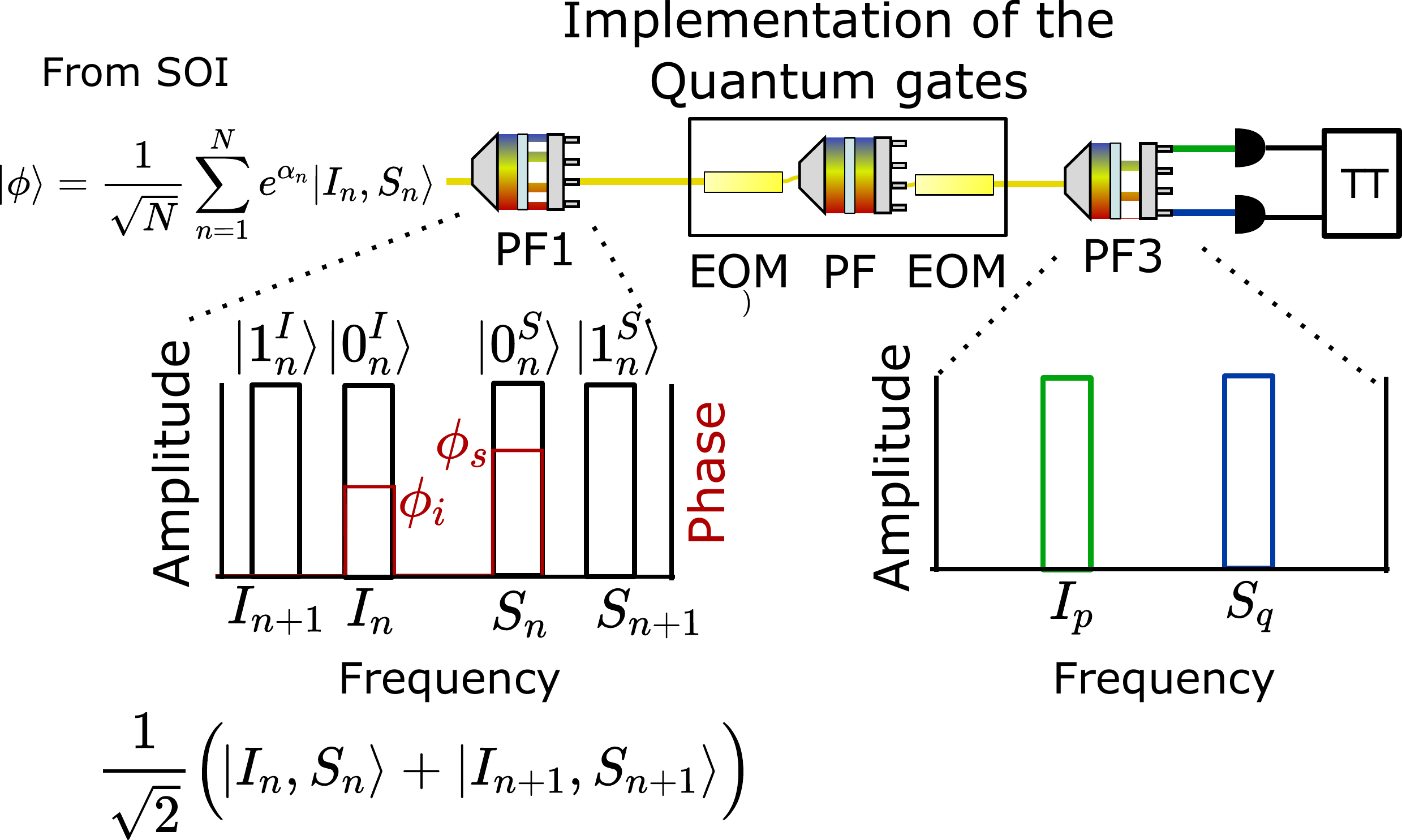}
\caption{Set up for the quantum state tomography. PF: Programmable filters, EOM: Electro-optic phase modulators. Insets are the action of the programmable filters on the frequency modes.
PF1 is used both as an amplitude filter to \IZ{select the four modes of the two qubits}, 
%carve the two qubit modes
and as a phase gate implementing a phase $\phi_{i}$ and $\phi_{s}$ on the frequency modes $I_n$ and $S_n$. The boxed device implement Identity or Hadamard gate on the qubits. All the projections required for the tomography are accessible with these two gates. PF3
\IZ{selects two modes $I_p$ and $S_q$, where p,q $\in$\{n,n+1\}}.}
\label{QST}
\end{figure}

\IZ{In this section, we demonstrate quantum state tomography of a frequency 2-dimensional maximally entangled state produced by the 21 GHz SOI micro-resonator. 
A first programmable filter PF1 (see Fig. \ref{QST}) selects two adjacent signal and idler mode pairs from the high-dimensional two-photon state of Eq. \ref{state} coming out of the
resonator to produce the biphoton maximally entangled state $\frac{1}{\sqrt{2}}(\ket{I_{n}S_{n}}+\ket{I_{n+1}S_{n+1}})$.

To perform the tomography of this state, single qubit rotations in the frequency domain are necessary.} We implement those using EOMs and PFs as demonstrated in \cite{lu_fully_2020} (see Appendix. \ref{quantum gates}). The setup is shown in Fig.\ref{QST} and uses these devices sequentially in a [EOM-PF-EOM] configuration.
\IZ{RF driving of the EOM at $\Omega$=FSR is possible here because the low FSR of our micro-resonator is compatible with the 40 GHz bandwidth of the EOM. With this configuration, the [EOM-PF-EOM] device shown on Fig. \ref{QST} can achieve parallel independent manipulation of signal and idler photons.}
%\AH{One great advantage of the frequency domain is the number of available quantum states. This number is limited by the bandwidth of the PF, and by the spacing allowed between the frequency modes.  
%The low FSR of our micro-resonator (21 GHz) is compatible with the 40 GHz bandwidth of the EOM. This allows us to drive the modulators with a frequency $\Omega=FSR$, and to parallelize two quantum gates on idler and signal photons. 

% The EOMs, driven by a RF signal at frequency $\Omega$, acts as a scatterer in the frequency domain, creating a superposition of frequency components separated by $\Omega$, weighed by the Bessel function \cite{capmany_quantum_2010}. To warranty the stability of the qubit in the frequency subspace, the frequency $\Omega$ must be equal to the spectral mode spacing of the qubit, here the FSR of the SOI micro-resonator. 
%  The programmable filter applies an arbitrary phase pattern where the phase for each frequency mode can be chosen independently. The [EOM-PF-EOM] combination creates controlled interferences, enabling the creation of a high-fidelity, tunable single qubit quantum operation such as Identity and Hadamard gates. 

%%%%
%Figure 
%%%

Fig.\ref{QST} shows a simplified setup for the quantum state tomography. In the full setup, polarizers are present before each EOM. Before each detector, a polarizer is also added to control the input polarization and a fibered notch filter around 900 nm is placed to suppress parasitic calibration light coming from the PF. 
%A first programmable filter (PF1) carves the high-dimensional two-photon state coming out of the resonator into two adjacent signal and idler mode pairs of the resonator to prepare the two-qubit-state: $\ket{S_{n}}$, $\ket{S_{n+1}}$ and $\ket{I_{n}}$, $\ket{I_{n+1}}$.

\IZ{Previously reported tomography in the spectral domain used a single EOM \cite{imany_50-ghz-spaced_2018,lu_full_2021, kues_-chip_2017,clementi_programmable_2023}. Such a simple setup can be used when the frequency spacing between two modes is too large to match the RF driving frequency $\Omega$ and then corresponds to a multiple of $\Omega$. This setup has the advantage of low loss but does not allow independent unitary operations on parallel qubits. }
%Quantum state tomography in the spectral domain has been performed on frequency domain quantum states using a single modulator \cite{imany_50-ghz-spaced_2018,lu_full_2021, kues_-chip_2017,clementi_programmable_2023}. The use of a single electro-optic modulator (EOM) instead of two EOMs and one phase modulator (PF) greatly reduces the insertion loss, allowing for a higher count number. It also enables the manipulation of higher-dimensional quantum states and a higher frequency spacing, as the spacing does not need to be equal to the RF frequency as the spacing can be a multiple of the RF driving frequency. In \cite{imany_50-ghz-spaced_2018}, the frequency spacing was equal to 50 GHz, with modulators driven at 25 GHz. However, this approach limits the ability to independently implement operations on parallel qubits using the same device. 
In \cite{lu_quantum_2018}, quantum state tomography was performed using two parallel tunable quantum gate with the same [EOM-PF-EOM] configuration.
\IZ{A Bayesian method based on measurements only in the $\mathbb{Z}$ and $\mathbb{X}$ bases was used to reconstruct the density matrix and a fidelity of 0.92$\pm$0.01 to a $\ket{\Psi^+}$ entangled state was obtained.}
%To reconstruct the density matrix, they use a Bayesian method based on the measurement in the $\mathbb{Z}$ and $\mathbb{X}$ Basis. and manage to get a Fidelity to an entangled state of 0.92$\pm$0.01 to a $\ket{\Psi^+}$. entangled state.}

Our logical qubits are defined as follows: $\ket{0^X_n} = \ket{X_n} $ and $\ket{1^X_n} = \ket{X_{n+1}}$, where $X=S,\,I$ refers to the logical signal or idler qubit (see Fig. \ref{QST}).
To perform the quantum state tomography, we need to project the two qubits onto four state vectors belonging to three different bases  $\mathbb{Z} = \left\{\ket{0^X_n},\ket{1^X_n} \right\}$, $\mathbb{X} = \left\{\ket{+^X_n},\ket{-^X_n} \right\}$ and $\mathbb{Y} = \left\{\ket{+i^X_n},\ket{-i^X_n} \right\}$,  where $\ket{\pm^X_n}= \frac{1}{\sqrt{2}} \left( \ket{0^X_n}\pm\ket{1^X_n}\right)$ and $\ket{\pm i^X_n}= \frac{1}{\sqrt{2}} \left( \ket{0^X_n}\pm i\ket{1^X_n}\right)$.
The quantum gate (see Fig. \ref{QST}) allows us to choose between the $\mathbb{Z}$ and $\mathbb{X}$ bases. To access the $\mathbb{Y}$ basis, we use PF1 to apply a relative phase shift $\phi_i$ ($\phi_s$) between the modes of the idler (signal) qubit. $C_{a,b}$ denotes the coincidence numbers corresponding to the projections on vectors $\ket{a}$ and $\ket{b}$ from the three different bases $\mathbb{X}$, $\mathbb{Y}$ and $\mathbb{Z}$. Table \ref{tab:counts_tomography} shows the recorded coincidences for the 16 projections performed on the two-qubit state  for n = 34.

\begin{figure}[ht]
\centering\includegraphics[width=10cm]{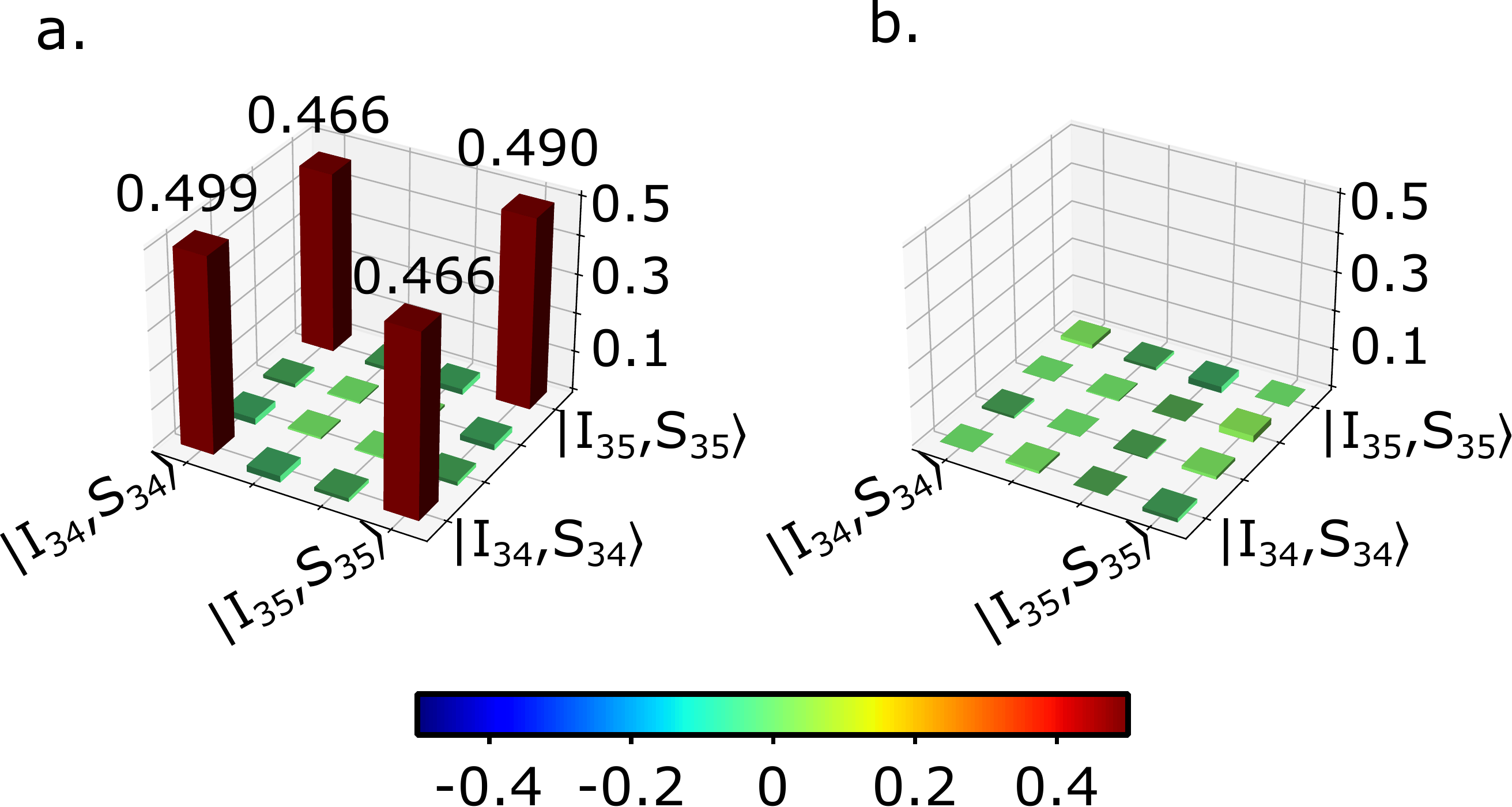}
\caption{Numerical reconstruction of the experimental density matrix of a two-qubit frequency-bin entangled state generated by the SOI resonator + PF1. a.: real part, and b.: imaginary part.  }
\label{exp_tomography}
\end{figure}

\begin{table}
    \centering
    \begin{tabular}{|c|c||c|c|}
    \hline
        Projections & Coinc. & Projections & Coinc.  \\
        \hline
        \hline
          $C_{0,0}$ & 1548 & $C_{+,0}$ & 716 \\
         $C_{0,1}$ & 36 & $C_{+,1}$ & 767 \\
         $C_{0,+}$ & 622 & $C_{+,+}$ & 1275 \\
         $C_{0,+i}$ & 663 &  $C_{+,+i}$ & 608 \\
         \hline
          $C_{1,0}$ & 22 & $C_{+i,0}$ & 837 \\
         $C_{1,1}$ & 1553 & $C_{+i,1}$ & 695 \\
         $C_{1,+}$ & 692 & $C_{+i,+}$ & 723 \\
         $C_{1,+i}$ & 664 & $C_{+i,+i}$ & 42 \\
         \hline

    \end{tabular}
    \caption{Coincidences for the two-photon projections $C_{a,b}$ integrated for 125 seconds, in a coincidence window of 1 ns.}
    \label{tab:counts_tomography}
\end{table}
Fig. \ref{exp_tomography} shows the reconstructed density matrix. Comparing it to the density matrix of a maximally entangled state $\ket{\phi^ +}= \frac{1}{\sqrt{2}}\left( \ket{I_{34},S_{34}} + \ket{I_{35},S_{35}}\right) $ we obtain a fidelity $\mathcal{F}=0.961\pm 0.007$. The errors are calculated using Monte-Carlo methods.

Having checked the fidelity of the produced frequency-entangled state, we take advantage of the broad bandwidth of our source and the versatility of the [EOM-PF-EOM] setup to parallelize the measurement of such high quality entangled states by performing 34 quantum gates on 17 frequency-entangled qubit pairs, \AH{in a first implementation of frequency-encoded quantum communication protocol.}

%%%%%%%%%%%%%%%%%%%%%%%%%%%%%%%%%%%%%%%%%%%%%%%%%%%%%%%%%%%%%%%%%%%%%%%%%%%%%%%%%%%%%%%%%%%%%%
%%%%%%%%%%%%%%%%%%%%%%%%%%%%%%%%%%%%%%%%%%%%%%%%%%%%%%%%%%%%%%%%%%%%%%%%%%%%%%%%%%%%%%%%%%%%%%
\section{Frequency-bin entangled photons for fully connected networks}

% Frequency-coded quantum protocols were investigated as early as 2007 in proof of concept studies with attenuated lasers and electro-optic phase modulator \cite{bloch_frequency-coded_2007}. This was followed by \cite{olislager_frequency-bin_2010}, which used frequency-entangled photons generated from a PPLN crystal to measure two-photon correlations and deduce Bell parameters. Here, we propose the first implementation of a fully-connected network using frequency-entangled photons.

In this section, we show proof-of-principle of a fully connected network of up to 5 users in which every user can share a secure key with every other user, using frequency-bin entangled qubits generated by our broadband photon pair source.
\IZ{In previous implementations or proof-of-principle experiments based on time-bin or polarization-based entangled sources \cite{joshi_trusted-node-free_2020,appas_flexible_2021-1,fitzke_scalable_2022,wengerowsky_entanglement-based_2018}, the main challenge was phase or polarization stabilization. In addition to previously mentionned assets, the frequency degree of freedom that we harness here has the advantage of requiring no phase stabilization for superposition analysis.}
 
%\AH{Compared to time-bin entanglement, using the frequency degree of freedom has the advantage of not requiring phase stabilization for superposition analysis. With polarization encoding, maintaining polarization stability over long distances and time is challenging. This difficulty can be addressed using frequency encoding. 
In this setup, polarization control is nevertheless necessary when entering the modulators. Recent advances have proposed schemes using polarization diversity electro-optic modulators (EOMs) to address this challenge \cite{sandoval_polarization_2019}, which is a relevant and additional asset for real world applications. 
 As we encode frequency qubits on distinct pairs of adjacent frequencies, one can use a PF \IZ{as a demultiplexer} to spectrally separate and distribute the pairs to respective users to create a network.
For each mode pair, compensation for the bi-photon residual phase is \IZ{required} in order to produce the desired $\ket{\phi^+}$ states \cite{lu_full_2021}.

To validate our scheme, we first show that the parallelization of the quantum gate enables measurement of entanglement for qubits separated by two guards modes, allowing for a measured crosstalk $\leq10^{-3}$ between adjacent quantum gates  (see appendix \ref{quantum gates}). Fig. \ref{parallel_tomography} shows the fidelity for the accessible frequency-bin entangled pairs. We start from the $n=10$ resonance from the pump frequency, to avoid any residual pump laser leakage.
The fidelity is higher than $0.8$ for 14 pairs. The few lower values are related to the dips observed in the coincidence spectrum shown in Fig. \ref{fig:JSI} around the 50th and 70th resonances from the pump.
%%%%
%Figure 
%%%
\begin{figure}[h!]
\centering\includegraphics[width=7cm]{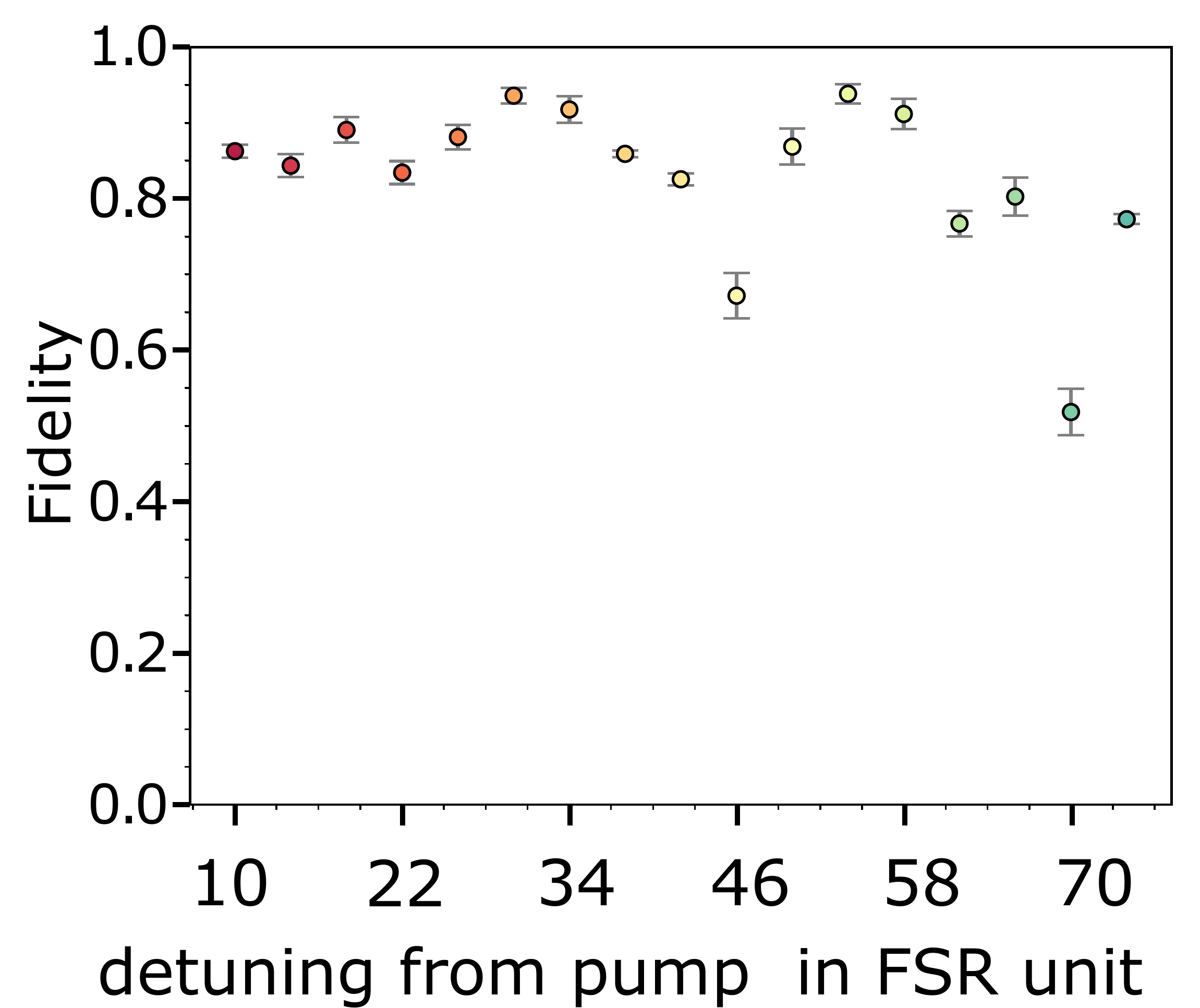}
\caption{Fidelity to a maximally entangled state for several frequency-bin entangled photon pairs. The x axis correspond to the distance of the frequency qubit to the pump frequency }
\label{parallel_tomography}
\end{figure}
%%%
% End figure
%%%

Important metrics for quantum networks are the key rate and qubit error rate. We deduce these parameters from our coincidence measurements, using the method proposed in \cite{autebert_multi-user_2016}. 
We use the coincidences in the $\mathbb{Z}$ basis ($C_{0,0}$, $C_{0,1}$, $C_{1,0}$, $C_{1,1}$) and the $\mathbb{X}$ basis ($C_{+,+}$, $C_{+,-}$, $C_{-,+}$ and  $C_{-,-}$) to compute the raw coincidence rate, QBER and sifted key rate (see appendix \ref{QKD_metrics}).

\begin{figure}[h!]
\centering\includegraphics[width=10cm]{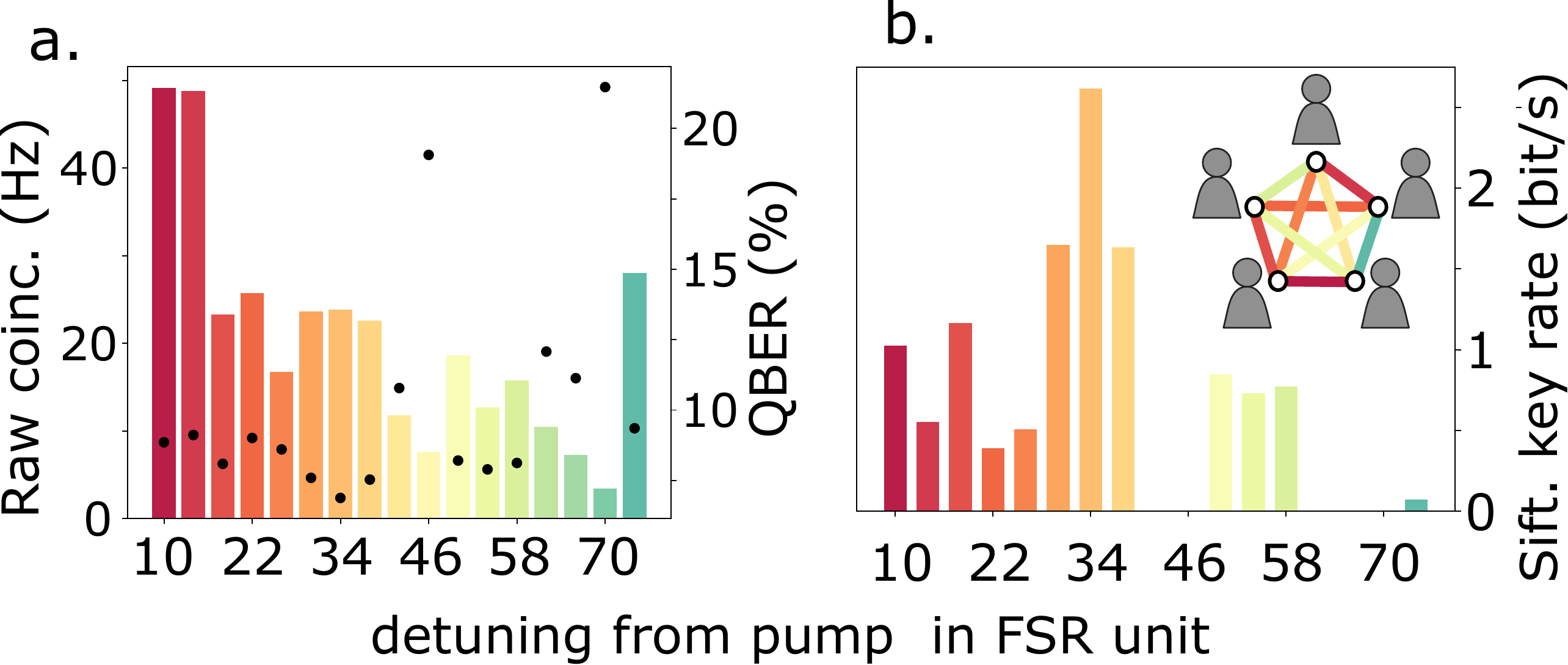}
\caption{ a.: Raw coincidences (bars) and qubit error rate (QBER) (dots) between two users, and b.: sifted key rate, calculated using the method in \cite{appas_flexible_2021-1} as a function of $n$, spectral detuning from the pump }
\label{raw_coinc_sifted}
\end{figure}

Fig. \ref{raw_coinc_sifted}.a shows, for each accessible pair, the number of raw coincidences $R_{raw}$ and the qubit error rate QBER. By comparing it with Fig. \ref{parallel_tomography}, we see that as expected, the qubits with a lower raw key rate (and higher QBER) are also the ones with the lower fidelity to an entangled state. 
A given pair achieves secure quantum communication only if the corresponding QBER is lower than the positive key rate threshold of 11\% \cite{lutkenhaus_security_2000}. 
Among the 17 accessible pairs, only 12 satisfy this condition. A sifted key rate in bit/s can also be derived from the raw rate $R_{raw}$ using the method in \cite{appas_flexible_2021-1} as depicted in Fig. \ref{raw_coinc_sifted}.b. 
The inset in Fig. \ref{raw_coinc_sifted}.b. shows how the distribution of 10 photon pairs is used to create a node-free quantum network of 5 users.

In this proof of principle experiment, the achieved key rates (0.5 to 2.5 bits/second) are limited by the performances of the source, as well as the insertion loss of the devices (14 dB for the 3 PFs and the 2 EOMs). Integrating these components on the same chip would help to limit losses and improve the overall insertion loss of the [EOM-PF-EOM] configuration to 73\%\cite{nussbaum_design_2022}. Several methods are proposed to achieve the integration of elements allowing the manipulation of frequency-bin qubits \cite{hu_-chip_2021,buddhiraju_arbitrary_2021}.
It is also worth noting that the number of accessible pairs could be increased by \IZ{setting} the pump frequency at the center of the PF bandwidth.

\section{Conclusion}
In this paper, we introduce a SOI micro-resonator for producing photon pairs at telecom wavelengths through SFWM and encoding frequency-bin qubits on adjacent pairs of resonances. Due to the broadband emission of the source and the 21 GHz FSR, we can generate up to 17 pairs of frequency entangled qubits.
We manipulate these qubits using quantum gates based on a single set of devices composed of one programmable filter between two electro-optic phase modulators, enabling independent control of each qubit. This leverages the ability to use these reprogrammable quantum gates for quantum information applications.

We demonstrate quantum state tomography with these quantum gates on parallel qubits, and assess a fidelity to a $\ket{\phi^+}$ entangled state of over 0.8 for 14 of them.

Finally, we demonstrate a local proof-of-concept of a fully connected network and compute the key rate and QBER of 17 photon pairs, taking into account the threshold for attacks on the quantum channels and error correction. We can distribute 10 photon pairs and create a fully-connected network of up to 5 users.

%leveraging the possibility to encode and distribute higher-dimensional quantum states.

%One limitation of this approach is the accumulated insertion loss of each device.
To scale up this approach, it is possible to encode and manipulate higher-dimensional quantum states with minimal additional resources using this setup, for example, by adding a single additional harmonic to the fundamental RF tone that drives the EOMs. Moreover, on-chip integration of electro-optic devices and programmable filters are a promising lead to reduce the losses (currently around 8.5 dB for our quantum gates implementation) \cite{nussbaum_design_2022}. On-chip integration has been proposed, for instance, using a lithium niobate resonator driven by RF signals \cite{hu_-chip_2021,buddhiraju_arbitrary_2021}.
\AH{Our work thus offers perspectives for scalable frequency-domain architectures for high-dimensional and resource-efficient quantum communications}

\section*{Ackowledgments}
This work has been supported by Region Ile-de-France in the framework of DIM SIRTEQ.
Lorenzo M. Procopio acknowledges the European
Union’s Horizon 2020 research under the Marie Skłodowska-Curie grant agreement No 800306.

\bibliography{template.bib}

\begin{thebibliography}{10}
\newcommand{\enquote}[1]{``#1''}

\bibitem{lu_quantum_2018}
H.-H. Lu, J.~M. Lukens, N.~A. Peters, B.~P. Williams, A.~M. Weiner, and
  P.~Lougovski, \enquote{Quantum interference and correlation control of
  frequency-bin qubits,} {\protect\JournalTitle{Optica}} \textbf{5}, 1455--1460
  (2018). [lu18].

\bibitem{imany_frequency-domain_2018}
P.~Imany, O.~D. Odele, M.~S. Alshaykh, H.-H. Lu, D.~E. Leaird, and A.~M.
  Weiner, \enquote{Frequency-domain {Hong}–{Ou}–{Mandel} interference with
  linear optics,} {\protect\JournalTitle{Optics Letters}} \textbf{43},
  2760--2763 (2018). [Imany18].

\bibitem{ponce_unlocking_2022}
M.~C. Ponce, A.~L.~M. Muniz, M.~Huber, and F.~Steinlechner, \enquote{Unlocking
  the frequency domain for high-dimensional quantum information processing,}
  (2022). ArXiv:2206.00969 [quant-ph].

\bibitem{khodadad_kashi_spectral_2021}
A.~Khodadad~Kashi and M.~Kues, \enquote{Spectral {Hong}–{Ou}–{Mandel}
  {Interference} between {Independently} {Generated} {Single} {Photons} for
  {Scalable} {Frequency}-{Domain} {Quantum} {Processing},}
  {\protect\JournalTitle{Laser \& Photonics Reviews}} \textbf{15}, 2000464
  (2021). [Khodadad21].

\bibitem{kues_-chip_2017}
M.~Kues, C.~Reimer, P.~Roztocki, L.~R. Cortés, S.~Sciara, B.~Wetzel, Y.~Zhang,
  A.~Cino, S.~T. Chu, B.~E. Little, D.~J. Moss, L.~Caspani, J.~Azaña, and
  R.~Morandotti, \enquote{On-chip generation of high-dimensional entangled
  quantum states and their coherent control,} {\protect\JournalTitle{Nature}}
  \textbf{546}, 622--626 (2017).

\bibitem{lu_full_2021}
H.-H. Lu, K.~V. Myilswamy, R.~S. Bennink, S.~Seshadri, M.~S. Alshaykh, J.~Liu,
  T.~J. Kippenberg, D.~E. Leaird, A.~M. Weiner, and J.~M. Lukens, \enquote{Full
  quantum state tomography of high-dimensional on-chip biphoton frequency combs
  with randomized measurements,} {\protect\JournalTitle{arXiv:2108.04124
  [physics, physics:quant-ph]}}  (2021). ArXiv: 2108.04124.

\bibitem{clementi_programmable_2023}
M.~Clementi, F.~A. Sabattoli, M.~Borghi, L.~Gianini, N.~Tagliavacche,
  H.~El~Dirani, L.~Youssef, N.~Bergamasco, C.~Petit-Etienne, E.~Pargon, J.~E.
  Sipe, M.~Liscidini, C.~Sciancalepore, M.~Galli, and D.~Bajoni,
  \enquote{Programmable frequency-bin quantum states in a nano-engineered
  silicon device,} {\protect\JournalTitle{Nature Communications}} \textbf{14},
  176 (2023).

\bibitem{borghi_reconfigurable_2023}
M.~Borghi, N.~Tagliavacche, F.~A. Sabattoli, H.~E. Dirani, L.~Youssef,
  C.~Petit-Etienne, E.~Pargon, J.~E. Sipe, M.~Liscidini, C.~Sciancalepore,
  M.~Galli, and D.~Bajoni, \enquote{A reconfigurable silicon photonics chip for
  the generation of frequency bin entangled qudits,}  (2023). ArXiv:2301.08475
  [quant-ph].

\bibitem{wang_integrated_2020}
J.~Wang, F.~Sciarrino, A.~Laing, and M.~G. Thompson, \enquote{Integrated
  photonic quantum technologies,} {\protect\JournalTitle{Nature Photonics}}
  \textbf{14}, 273--284 (2020).

\bibitem{zhang_correlated_2016}
X.~Zhang, Y.~Zhang, C.~Xiong, and B.~J. Eggleton, \enquote{Correlated photon
  pair generation in low-loss double-stripe silicon nitride waveguides,}
  {\protect\JournalTitle{Journal of Optics}} \textbf{18}, 074016 (2016).

\bibitem{samara_high-rate_2019}
F.~Samara, A.~Martin, C.~Autebert, M.~Karpov, T.~J. Kippenberg, H.~Zbinden, and
  R.~Thew, \enquote{High-rate photon pairs and sequential {Time}-{Bin}
  entanglement with {Si} $_{\textrm{3}}$ {N} $_{\textrm{4}}$ microring
  resonators,} {\protect\JournalTitle{Optics Express}} \textbf{27}, 19309
  (2019).

\bibitem{yin_frequency_2021}
Z.~Yin, K.~Sugiura, H.~Takashima, R.~Okamoto, F.~Qiu, S.~Yokoyama, and
  S.~Takeuchi, \enquote{Frequency correlated photon generation at telecom band
  using silicon nitride ring cavities,} {\protect\JournalTitle{Optics Express}}
  \textbf{29}, 4821 (2021).

\bibitem{clemmen_continuous_2009}
S.~Clemmen, K.~P. Huy, W.~Bogaerts, R.~G. Baets, P.~Emplit, and S.~Massar,
  \enquote{Continuous wave photon pair generation in silicon-on-insulator
  waveguides and ring resonators,} {\protect\JournalTitle{Optics Express}}
  \textbf{17}, 16558 (2009).

\bibitem{kobayashi_frequency-domain_2016}
T.~Kobayashi, R.~Ikuta, S.~Yasui, S.~Miki, T.~Yamashita, H.~Terai, T.~Yamamoto,
  M.~Koashi, and N.~Imoto, \enquote{Frequency-domain {Hong}–{Ou}–{Mandel}
  interference,} {\protect\JournalTitle{Nature Photonics}} \textbf{10},
  441--444 (2016).

\bibitem{raymer_interference_2010}
M.~G. Raymer, S.~J. van Enk, C.~J. McKinstrie, and H.~J. McGuinness,
  \enquote{Interference of two photons of different color,}
  {\protect\JournalTitle{Optics Communications}} \textbf{283}, 747--752 (2010).

\bibitem{olislager_creating_2014}
L.~Olislager, E.~Woodhead, K.~Phan~Huy, J.-M. Merolla, P.~Emplit, and
  S.~Massar, \enquote{Creating and manipulating entangled optical qubits in the
  frequency domain,} {\protect\JournalTitle{Physical Review A}} \textbf{89},
  052323 (2014).

\bibitem{olislager_frequency-bin_2010}
L.~Olislager, J.~Cussey, A.~T. Nguyen, P.~Emplit, S.~Massar, J.-M. Merolla, and
  K.~P. Huy, \enquote{Frequency-bin entangled photons,}
  {\protect\JournalTitle{Physical Review A}} \textbf{82}, 013804 (2010).

\bibitem{bloch_frequency-coded_2007}
M.~Bloch, S.~W. McLaughlin, J.-M. Merolla, and F.~Patois,
  \enquote{Frequency-coded quantum key distribution,}
  {\protect\JournalTitle{Optics Letters}} \textbf{32}, 301 (2007).

\bibitem{lukens_frequency-encoded_2017}
J.~M. Lukens and P.~Lougovski, \enquote{Frequency-encoded photonic qubits for
  scalable quantum information processing,} {\protect\JournalTitle{Optica}}
  \textbf{4}, 8--16 (2017).

\bibitem{lu_electro-optic_2018}
H.-H. Lu, J.~M. Lukens, N.~A. Peters, O.~D. Odele, D.~E. Leaird, A.~M. Weiner,
  and P.~Lougovski, \enquote{Electro-{Optic} {Frequency} {Beamsplitters} and
  {Tritters} for {High}-{Fidelity} {Photonic} {Quantum} {Information}
  {Processing},} {\protect\JournalTitle{Physical Review Letters}} \textbf{120},
  030502 (2018). ArXiv: 1712.03992.

\bibitem{lu_fully_2020}
H.-H. Lu, E.~M. Simmerman, P.~Lougovski, A.~M. Weiner, and J.~M. Lukens,
  \enquote{Fully {Arbitrary} {Control} of {Frequency}-{Bin} {Qubits},}
  {\protect\JournalTitle{Physical Review Letters}} \textbf{125}, 120503 (2020).

\bibitem{lu_controlled-not_2019}
H.-H. Lu, J.~M. Lukens, B.~P. Williams, P.~Imany, N.~A. Peters, A.~M. Weiner,
  and P.~Lougovski, \enquote{A controlled-{NOT} gate for frequency-bin qubits,}
  {\protect\JournalTitle{npj Quantum Information}} \textbf{5}, 1--8 (2019).

\bibitem{lu_subatomic_2019}
H.-H. Lu, N.~Klco, J.~M. Lukens, T.~D. Morris, T.~D. Morris, A.~Bansal,
  A.~Ekström, G.~Hagen, G.~Hagen, T.~Papenbrock, T.~Papenbrock, A.~M. Weiner,
  M.~J. Savage, and P.~Lougovski, \enquote{Subatomic {Many}-{Body} {Physics}
  {Simulations} on a {Quantum} {Frequency} {Processor},} in \emph{Conference on
  {Lasers} and {Electro}-{Optics} (2019), paper {FTh3A}.6,}  (Optical Society
  of America, 2019), p. FTh3A.6.

\bibitem{lu_high-dimensional_2022}
H.-H. Lu, N.~B. Lingaraju, D.~E. Leaird, A.~M. Weiner, and J.~M. Lukens,
  \enquote{High-dimensional discrete {Fourier} transform gates with a quantum
  frequency processor,} {\protect\JournalTitle{Optics Express}} \textbf{30},
  10126 (2022).

\bibitem{imany_50-ghz-spaced_2018}
P.~Imany, J.~A. Jaramillo-Villegas, O.~D. Odele, K.~Han, D.~E. Leaird, J.~M.
  Lukens, P.~Lougovski, M.~Qi, and A.~M. Weiner, \enquote{50-{GHz}-spaced comb
  of high-dimensional frequency-bin entangled photons from an on-chip silicon
  nitride microresonator,} {\protect\JournalTitle{Optics Express}} \textbf{26},
  1825--1840 (2018).

\bibitem{Boeuf:16}
F.~Boeuf, S.~Cr\'{e}mer, E.~Temporiti, M.~Fer\`{e}, M.~Shaw, C.~Baudot,
  N.~Vulliet, T.~Pinguet, A.~Mekis, G.~Masini, H.~Petiton, P.~L. Maitre,
  M.~Traldi, and L.~Maggi, \enquote{Silicon photonics r\&d and manufacturing on
  300-mm wafer platform,} {\protect\JournalTitle{J. Lightwave Technol.}}
  \textbf{34}, 286--295 (2016).

\bibitem{do_wideband_2020}
P.~T. Do, C.~Alonso-Ramos, X.~Le~Roux, I.~Ledoux, B.~Journet, and E.~Cassan,
  \enquote{Wideband tunable microwave signal generation in a
  silicon-micro-ring-based optoelectronic oscillator,}
  {\protect\JournalTitle{Scientific Reports}} \textbf{10}, 6982 (2020).

\bibitem{joshi_trusted-node-free_2020}
S.~K. Joshi, D.~Aktas, S.~Wengerowsky, M.~Lončarić, S.~P. Neumann, B.~Liu,
  T.~Scheidl, G.~C. Lorenzo, Z.~Samec, L.~Kling, A.~Qiu, M.~Razavi,
  M.~Stipčević, J.~G. Rarity, and R.~Ursin, \enquote{A trusted-node-free
  eight-user metropolitan quantum communication network,}
  {\protect\JournalTitle{Science Advances}} \textbf{6}, eaba0959 (2020). ArXiv:
  1907.08229.

\bibitem{appas_flexible_2021-1}
F.~Appas, F.~Baboux, M.~I. Amanti, A.~Lemaítre, F.~Boitier, E.~Diamanti, and
  S.~Ducci, \enquote{Flexible entanglement-distribution network with an
  {AlGaAs} chip for secure communications,} {\protect\JournalTitle{npj Quantum
  Information}} \textbf{7}, 1--10 (2021).

\bibitem{fitzke_scalable_2022}
E.~Fitzke, L.~Bialowons, T.~Dolejsky, M.~Tippmann, O.~Nikiforov, T.~Walther,
  F.~Wissel, and M.~Gunkel, \enquote{Scalable {Network} for {Simultaneous}
  {Pairwise} {Quantum} {Key} {Distribution} via {Entanglement}-{Based}
  {Time}-{Bin} {Coding},} {\protect\JournalTitle{PRX Quantum}} \textbf{3},
  020341 (2022).

\bibitem{wengerowsky_entanglement-based_2018}
S.~Wengerowsky, S.~K. Joshi, F.~Steinlechner, H.~Hübel, and R.~Ursin,
  \enquote{An entanglement-based wavelength-multiplexed quantum communication
  network,} {\protect\JournalTitle{Nature}} \textbf{564}, 225--228 (2018).

\bibitem{sandoval_polarization_2019}
O.~E. Sandoval, N.~B. Lingaraju, P.~Imany, D.~E. Leaird, M.~Brodsky, and A.~M.
  Weiner, \enquote{Polarization diversity phase modulator for measuring
  frequency-bin entanglement of a biphoton frequency comb in a depolarized
  channel,} {\protect\JournalTitle{Optics Letters}} \textbf{44}, 1674 (2019).

\bibitem{autebert_multi-user_2016}
C.~Autebert, J.~Trapateau, A.~Orieux, A.~Lemaître, C.~Gomez-Carbonell,
  E.~Diamanti, I.~Zaquine, and S.~Ducci, \enquote{Multi-user quantum key
  distribution with entangled photons from an {AlGaAs} chip,}
  {\protect\JournalTitle{Quantum Science and Technology}} \textbf{1}, 01LT02
  (2016).

\bibitem{lutkenhaus_security_2000}
N.~Lütkenhaus, \enquote{Security against individual attacks for realistic
  quantum key distribution,} {\protect\JournalTitle{Physical Review A}}
  \textbf{61}, 052304 (2000).

\bibitem{nussbaum_design_2022}
B.~E. Nussbaum, A.~J. Pizzimenti, N.~B. Lingaraju, H.-H. Lu, and J.~M. Lukens,
  \enquote{Design {Methodologies} for {Integrated} {Quantum} {Frequency}
  {Processors},}  (2022). ArXiv:2204.12320 [physics, physics:quant-ph].

\bibitem{hu_-chip_2021}
Y.~Hu, M.~Yu, D.~Zhu, N.~Sinclair, A.~Shams-Ansari, L.~Shao, J.~Holzgrafe,
  E.~Puma, M.~Zhang, and M.~Lončar, \enquote{On-chip electro-optic frequency
  shifters and beam splitters,} {\protect\JournalTitle{Nature}} \textbf{599},
  587--593 (2021).

\bibitem{buddhiraju_arbitrary_2021}
S.~Buddhiraju, A.~Dutt, M.~Minkov, I.~A.~D. Williamson, and S.~Fan,
  \enquote{Arbitrary linear transformations for photons in the frequency
  synthetic dimension,} {\protect\JournalTitle{Nature Communications}}
  \textbf{12}, 2401 (2021).

\end{thebibliography}
%\bibliography{OSA-template}

\appendix

%%%%%%%%%%%%%%%%%%%%%%%%%%%%%%%%%%%%
\section{Frequency-domain quantum gates}
\label{quantum gates}
\subsection{Gate characterization}

\begin{figure}[h!]
    \centering
    \includegraphics[width=8cm]{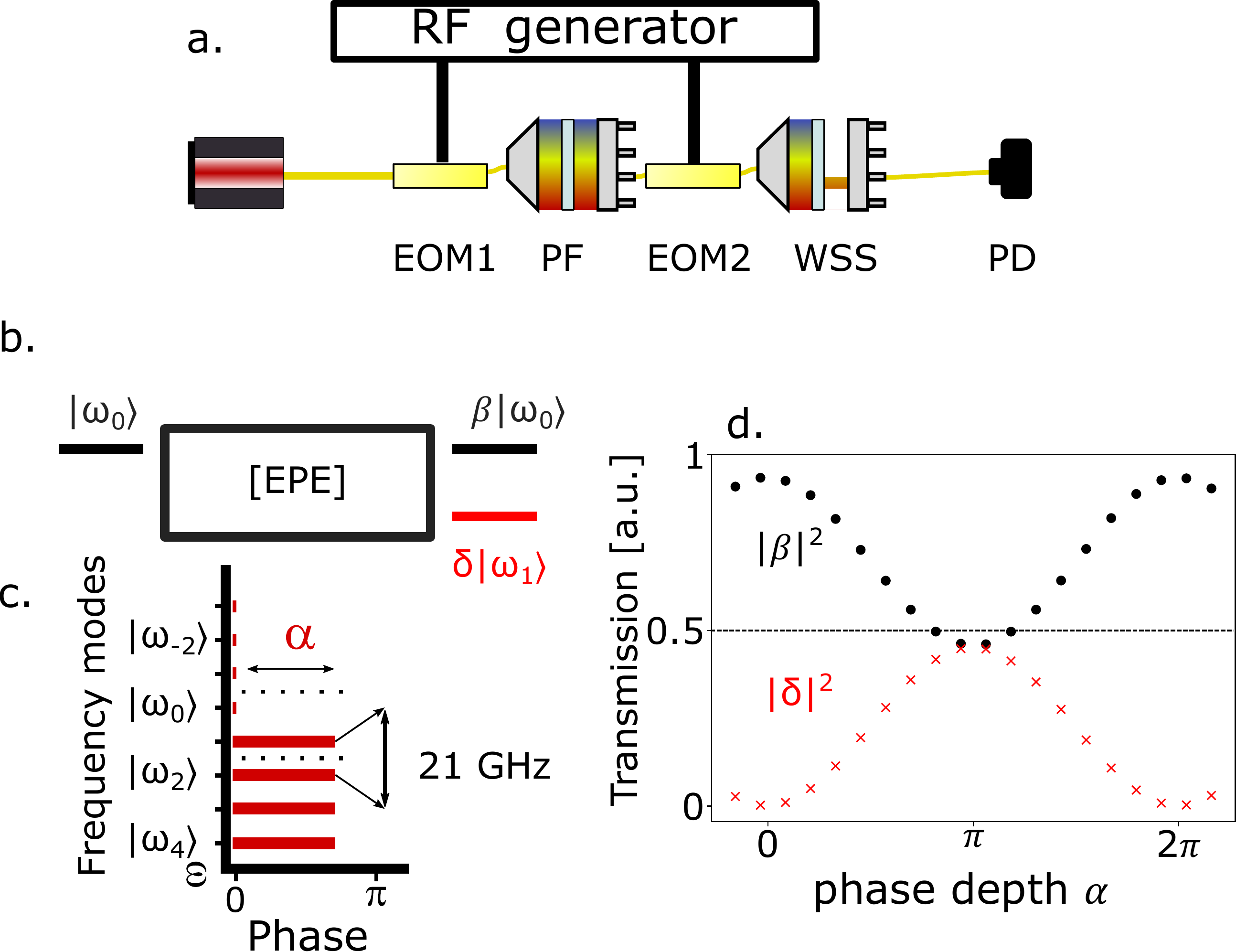}
    \caption{Classical characterization of the quantum gate. a. : %scheme of 
    principle of a frequency-domain operation. b. : principle of a frequency domain quantum gate. c: phase pattern applied by the PF %to tune the operation. 
    d. : measured tunability of the quantum gate operation
}
    \label{fig:classical_carac}
\end{figure}

In this section, we use a combination of EOMs, and PFs to create frequency-bin quantum gates\cite{lu_electro-optic_2018,lu_fully_2020}, that can be parallelized to a large number of qubits.
The setup in Fig. \ref{fig:classical_carac}.a. presents the quantum gate, composed of two electro-optic phase modulators iXblue MPZ-ln-40 (EOM1 and EOM2), driven by a multi-channel RF sine wave generator Anapico APMS33G, and one phase-only programmable filter Finisar WaveShaper 1000A (PF). The overall insertion loss of the quantum gate is around 9 dB from the input of the first modulator to the output of the second. The time dependent phase shift $\phi(t)$ 
%viewed by the optical field 
applied by the EOM to the optical wave is proportional to the sine wave produced by the RF generator : 
\begin{equation}
    \phi(t) = \mu\cos\left(\Omega t + \theta \right),
\end{equation}
where the RF frequency $\Omega$ %is the RF frequency
determines the mode separation between the generated frequency modes $\ket{\omega_n} = \ket{\omega_0 + n\Omega}$, and $\mu$ is the modulation index, proportional to the RF voltage V : $\mu = \pi\frac{V}{V_\pi}$. $V_\pi$ is the voltage %necessary 
giving rise to a $\pi$ phase shift. The specific settings of the modulators and programmable filter allowing to implement the quantum gate (tomography) are the following: 
both modulators must be driven with the modulation index $\mu =0.81$ and the relative phase between them must be set to $\pi$. This operation is achieved by using the multi-channel available on the generator, on which we can set a relative phase, ranging from 0 to $\pi$. It is noteworthy that compensation of the dispersion is required in order to keep this relative RF phase shift constant at all optical frequencies within the PF bandwidth.
Another PF is used as a wavelength selecting switch (WSS) after the quantum gate to select the frequency mode to be detected by the IR photo-diode (PD). %the intensity of the light in a specific frequency mode. The light then goes to a a IR photo-diode (PD). 
Fig. \ref{fig:classical_carac}.b. Shows the principle of the operation performed. We define a frequency-bin qubit as a qubit for which the information is encoded on two frequency modes, here generically called $\ket{\omega_0}$ and $\ket{\omega_1}$. The goal of the quantum gate, is apply a controlled rotation on the qubit, here represented by the coefficients $\beta$ and $\delta$. 
To characterize the quantum operation, we use the fidelity $\mathcal{F}$ and the success probability $\mathcal{P}$ \cite{lu_electro-optic_2018}: 
\begin{equation}
\mathcal{F} = \frac{Tr(W^\dagger T)Tr(T^\dagger W)}{Tr(W^\dagger W)Tr(T^\dagger T)}, \; \; \mathcal{P} = \frac{Tr(W^\dagger W)}{Tr(T^\dagger T)}
\end{equation}

Where the fidelity corresponds  to how close the realiezd operation ($W$) is to the ideal case ($T$), and the success probability measures the unitarity of the operation, accounting for the energy loss towards unwanted frequency modes.

 Fig. \ref{fig:classical_carac}.c. represents the phase pattern applied by the PF in order to tune the operation of the gate. A step phase of height $\alpha$ is applied between the two qubit modes.
In order to characterize the spectral processing implemented by this single qubit gate, we use coherent light as it is sufficient to characterize a single qubit gate.

 \ref{fig:classical_carac}d. shows the effect of the phase difference $\alpha$ applied by the PF between the two qubit modes on the quantum gate, evolving from Identity ($\alpha = 0$) to Hadamard operation ($\alpha = \pi$). The success probability experimentally reaches $\mathcal{P}=0,95$ implies that the intrinsic transmission of the quantum gate is slightly below 0.5 but the splitting between the two modes is balanced, leading to a Fidelity $\mathcal{P}=0,99$. The success probability does not take into account the insersion loss of each devices, around 8.5 dB (3.5 for the PF, and 2.5 for each modulators).

%\ref{fig:classical_carac}.d., we observe that the middle point, representing equal superposition, should have a maximum transmission of 0.5 for each modes. The residual loss arise from the energy being lost in adjacent neighboring modes.
\subsection{Gate parallelization}
In these experiments, the mode spacing is taken to be 21 GHz. This is compliant with the resolution of the programmable filters (10 GHz), and the accessible modulation bandwidth of commercial EOMs (40 GHz). The phase pattern depicted in Fig.  \ref{fig:classical_carac}.c. shows a step phase applied between the two modes of the qubit. By %periodically
applying a phase change over several pairs of frequency modes, it is possible to parallelize independent quantum operations over several frequency-bin qubits.

\begin{figure}
    \centering
    \includegraphics[width=8cm]{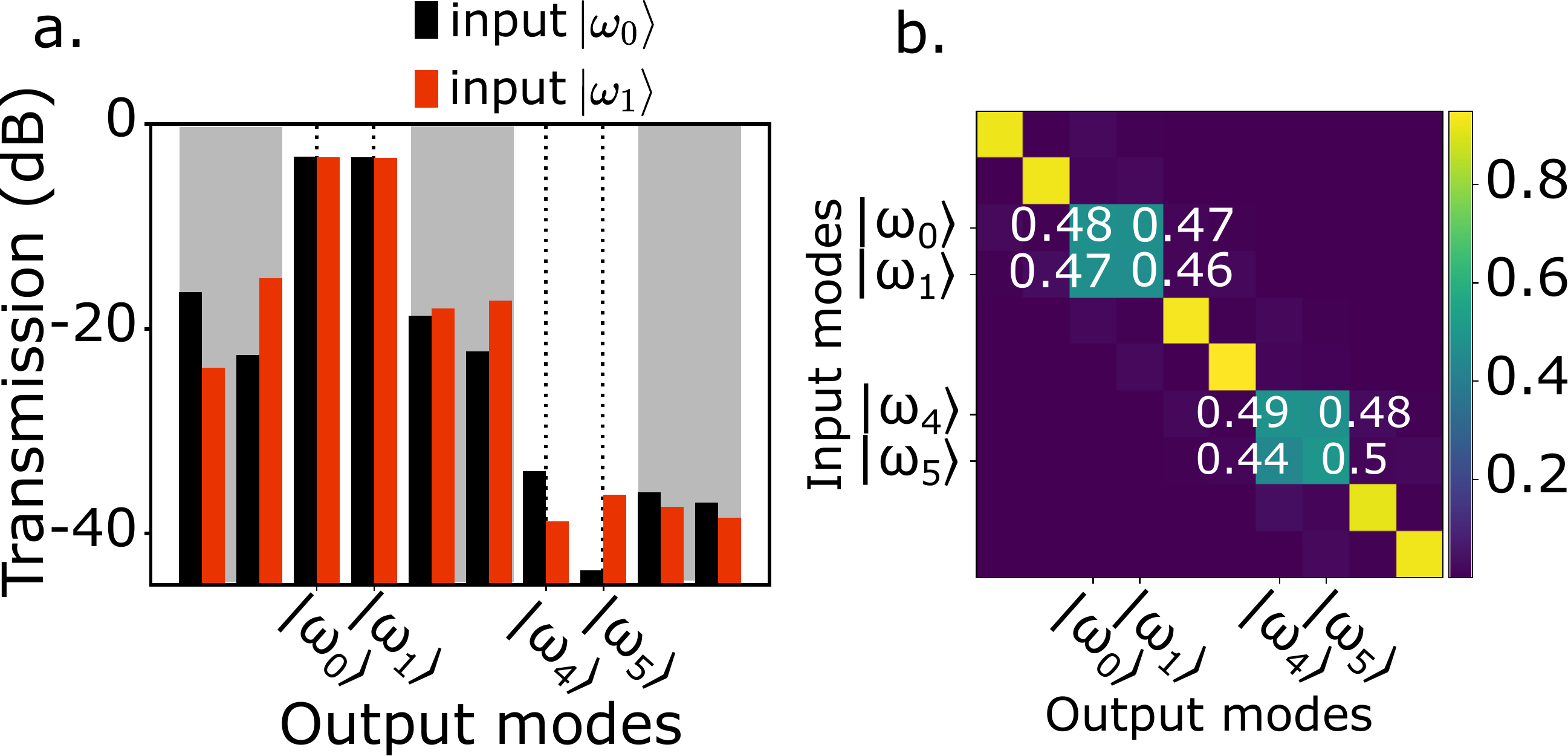}
    \caption{ a : light coupling from mode $\ket{\omega_0}$ (black) or $\ket{\omega_1}$ (red) into 9 neighboring modes. The grey shaded areas represent the guard modes that are not used for parallelization. b. Transmission measurement for 9  frequency modes when a Hadamard gate is applied to two qubits separated by two guard modes. }
    \label{fig:classical_parallelization}
\end{figure}

It is important, for parallelization, that two consecutive gates are isolated from each other. To determine what is the minimum desired spacing between each qubits, we send light through the quantum gate, tuned to realize the Hadamard transformation. We define the crosstalk between the qubit as the portion of light being converted from one qubit to another.
The experimental results are shown in Fig. \ref{fig:classical_parallelization}.a, where the black(red) lines correspond to the measured light, when we input light in the mode $\omega_0$($\omega_1$). 
Fig. 
From this figure, we see that two separation modes between consecutive operations are enough to ensure less than $10^{-3}$ crosstalk between two qubits. 
Fig.\ref{fig:classical_parallelization}.b. shows the intensity transmission matrix of a 10 dimensional frequency space, where we apply two Hadamard gates on the frequency modes $\omega_0,\,\omega_1$, and $\omega_4,\,\omega_5$. This leverages the possibility to operate in parallel on many qubits.
To allow the parallelization of the quantum gates for the whole C-band, it is necessary to compensate for the dispersion in the fiber between the two modulators. To this end, the PF in the middle of the [EOM-PF-EOM] setup applies a supplementary phase shift over the whole PF spectrum corresponding to a negative dispersion of -0.4 ps/nm.

%%%%%%%%%%
%%%%%%%%%%%%%
%%%%%%%%%%%
\section{Evaluating the performances for quantum key distribution}
\label{QKD_metrics}
%%%%%%%%%%%%%%
%the symbols '+' and '-' referring to the diagonal and anti-diagonal states \IZ{déjà introduit pour la tomographie}.
We use the coincidences in the $\mathbb{Z}$ basis ($C_{0,0}$, $C_{0,1}$, $C_{1,0}$, $C_{1,1}$) and the $\mathbb{X}$ basis ($C_{+,+}$, $C_{+,-}$, $C_{-,+}$ and  $C_{-,-}$) to compute the raw coincidence rate, QBER and sifted key rate. The raw numbers of coincidences in these bases are equal to \cite{appas_flexible_2021-1}
\begin{align}
    C_\mathbb{Z} = \frac{1}{2}\left(C_{0,0}+C_{0,1}+C_{1,0}+C_{1,1}    \right) \\
    C_\mathbb{X} = \frac{1}{2}\left(C_{+,+}+C_{+,-}+C_{-,+}+C_{-,-}    \right).
\end{align}

%We can compute the raw coincidence rate using $C_\mathbb{Z}$ and $C_\mathbb{X}$.
If the integration time is $\tau$, the raw coincidence rate is then
\begin{equation}
    R_{raw} = \frac{1}{2}\frac{\left( C_\mathbb{Z}+  C_\mathbb{X} \right)}{\tau}.
\end{equation}
The qubit error rate (QBER) is the ratio of accidental to total coincidences over the two bases
\begin{equation}
    e = \frac{C_{0,1}+ C_{1,0} + C_{+,-} + C_{-,+} }{C_\mathbb{Z} + C_\mathbb{X}}
\end{equation}

We compute the sifted key rate using formulas in \cite{autebert_multi-user_2016} accounting for error correction.

%%%%%%%%%%%%%%%%%%%%%%%%%%%%%%%%%%%%%%%%%%%%%

\end{document}